\documentclass[notitlepage,twocolumn,prl,tightenlines,nofootinbib,superscriptaddress]{revtex4}

\pdfoutput=1

\usepackage{amsmath}
\usepackage{amssymb,amsfonts}
\usepackage{graphicx}
\graphicspath{{figs/}{figs_lo/}}

\newcommand{\ie}{\textit{i.e.}}
\newcommand{\etal}{\textit{et~al.}}
\newcommand{\mathnotation}[2]{\newcommand{#1}{\ensuremath{#2}}}

\renewcommand{\time}{t}                
\mathnotation{\xc}{\theta}             
\mathnotation{\yc}{y}                  
\mathnotation{\Uc}{\Omega}             
\mathnotation{\A}{A}                   
\mathnotation{\xcp}{\bar\xc}           
\mathnotation{\ycp}{\bar\yc}           
\mathnotation{\T}{T}                   
\mathnotation{\xcfp}{\xc_*}            
\mathnotation{\ycfp}{\yc_*}            
\mathnotation{\decfp}{\mu}             
\mathnotation{\decvar}{\alpha}         
\mathnotation{\gap}{d}                 
\mathnotation{\gaps}{d_*}              
\mathnotation{\dotgap}{\skew{8}\dot{d}}
\mathnotation{\tin}{\time_{\mathrm{init}}}
\mathnotation{\tsep}{\time_{\mathrm{sep}}}

\begin{document}

\title{Rotation shields chaotic mixing regions from no-slip walls}

\author{E. Gouillart}
\affiliation{Surface du Verre et Interfaces, UMR 125 CNRS/Saint-Gobain,
  93303 Aubervilliers, France}
\author{J.-L. Thiffeault}
\affiliation{Department of Mathematics, University of Wisconsin --
  Madison, WI 53706, USA}
\author{O. Dauchot}
\affiliation{Service de Physique de l'Etat Condens\'e, DSM, CEA
  Saclay, URA2464, 91191 Gif-sur-Yvette Cedex, France}

\date{September 21, 2009}


\begin{abstract}
We report on the decay of a passive scalar in chaotic mixing protocols
where the wall of the vessel is rotated, or a net drift of fluid elements
near the wall is induced at each period. As a result the fluid domain
is divided into a central isolated chaotic region and a
peripheral regular region. Scalar patterns obtained in experiments and
simulations converge to a strange eigenmode and follow an exponential
decay. This contrasts with previous experiments [Gouillart~\etal,
\emph{Phys. Rev. Lett.} \textbf{99}, 114501 (2007)]
with a chaotic region spanning the whole domain, where
fixed walls constrained mixing to follow a slower algebraic decay.
Using a linear analysis of the flow close to the wall,
as well as numerical simulations of Lagrangian trajectories, we study the
influence of the rotation velocity of the wall on the size of the chaotic
region, the approach to its bounding separatrix, and the decay rate of
the scalar.
\end{abstract}

\maketitle 

Chaotic advection is often presented~\cite{Aref1984,Ottino} as an
efficient alternative to turbulence for the mixing of viscous fluids
or delicate substances, as often encountered for example in the food
and pharmaceutical industries. The repeated stretching and folding of
fluid particles, induced for instance by mechanical stirrers,
transforms an heterogeneity into a striated pattern with high
gradients of the concentration field. This greatly enhances molecular
diffusion, which accelerates the homogenization process.  In contrast
to turbulence, the velocity field that causes chaotic advection is
spatially smooth and coherent in time.  Fluid particles may therefore
reside for long times in regions where stretching is much lower than
average, before wandering to the rest of the domain.  The global
mixing rate is slowed down by such fluid particles whose motion is not
ergodic on the typical timescales of a mixing experiment.  For
example, ``sticky'' KAM islands trap particles in their vicinity for
long times~\cite{Pikovsky2003}.  Recent theory~\cite{groupedtheory},
numerical simulations~\cite{groupednumerics}, and
experiments~\cite{Gouillart2007,Gouillart2008,Gouillart2009} have
shown that the vicinity of a no-slip vessel wall is a region of poor
stretching that can slow down the whole mixing process.

\begin{figure}[b!]
\vspace{-0.3cm}
\begin{center}
\includegraphics[width=0.47\columnwidth, height=0.47\columnwidth]{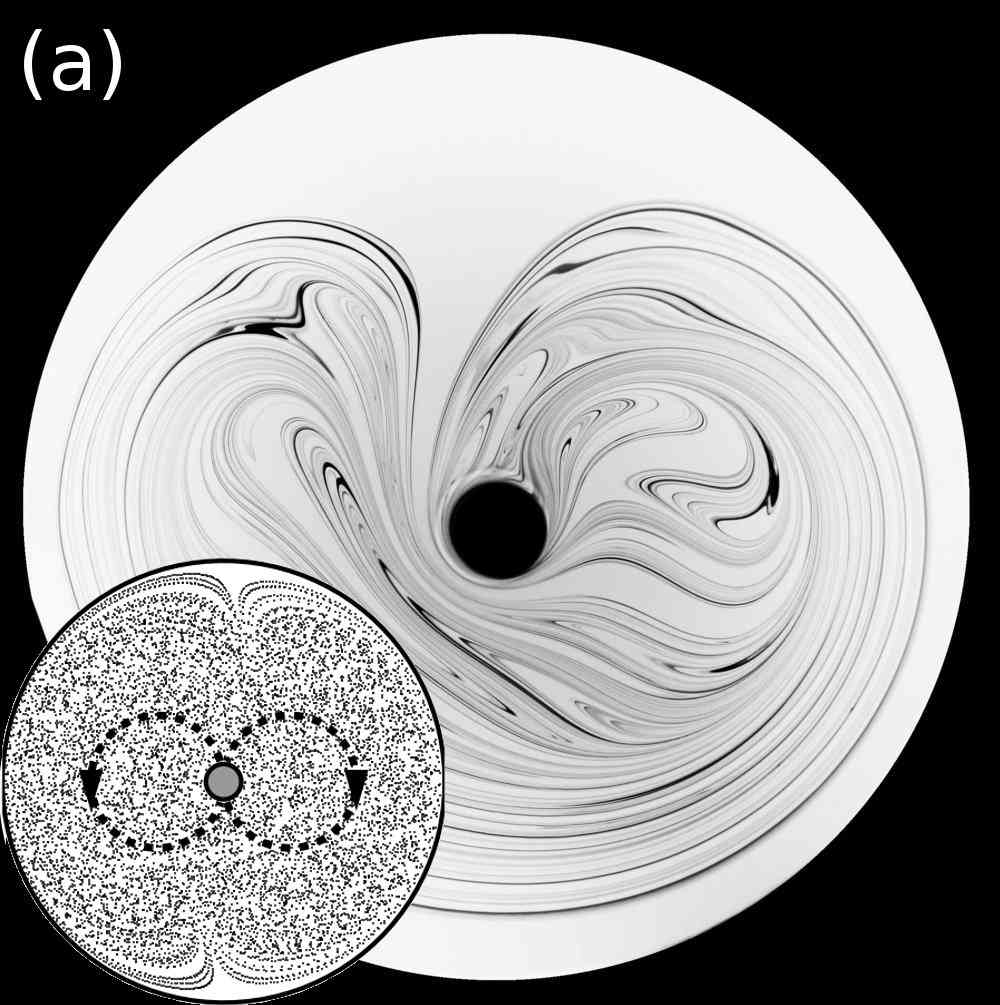}
\includegraphics[width=0.47\columnwidth, height=0.47\columnwidth]{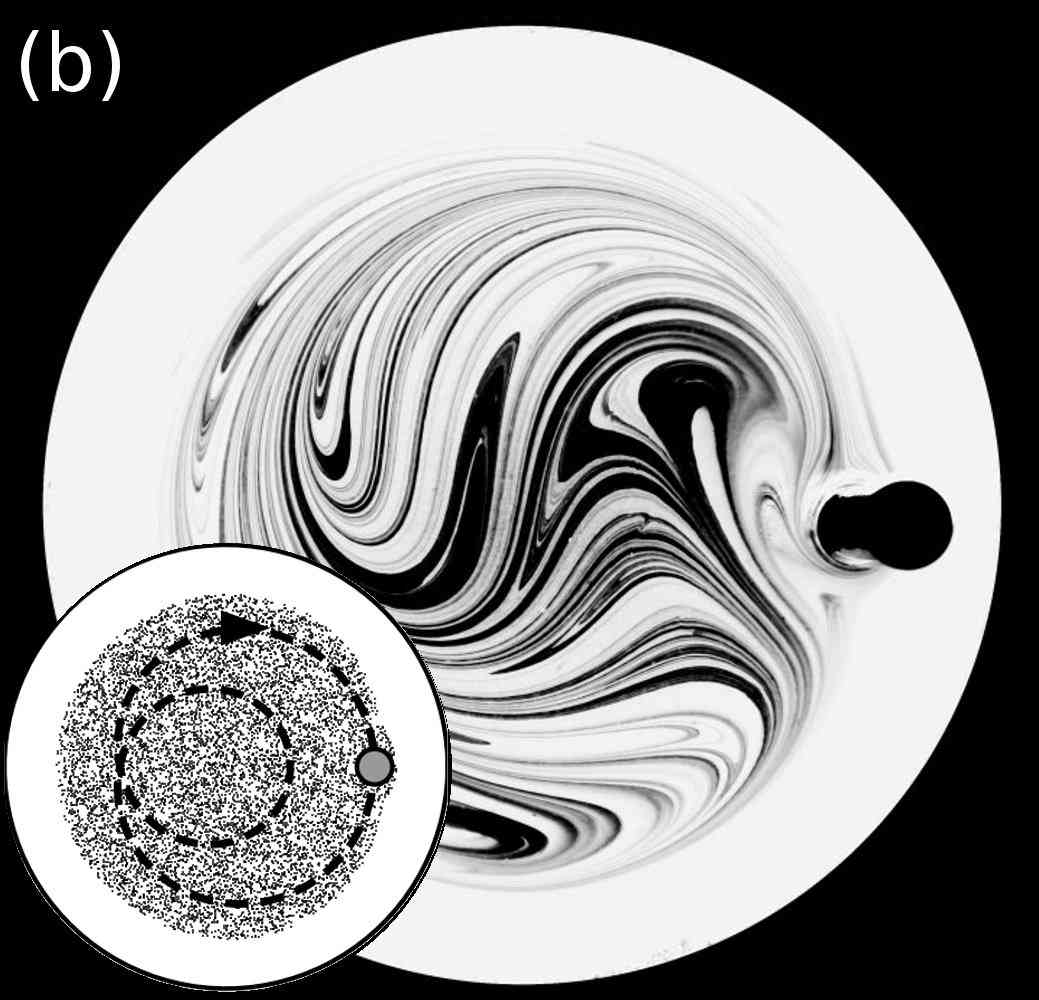}
\vskip 0.5em
\includegraphics[width=0.47\columnwidth, height=0.47\columnwidth]{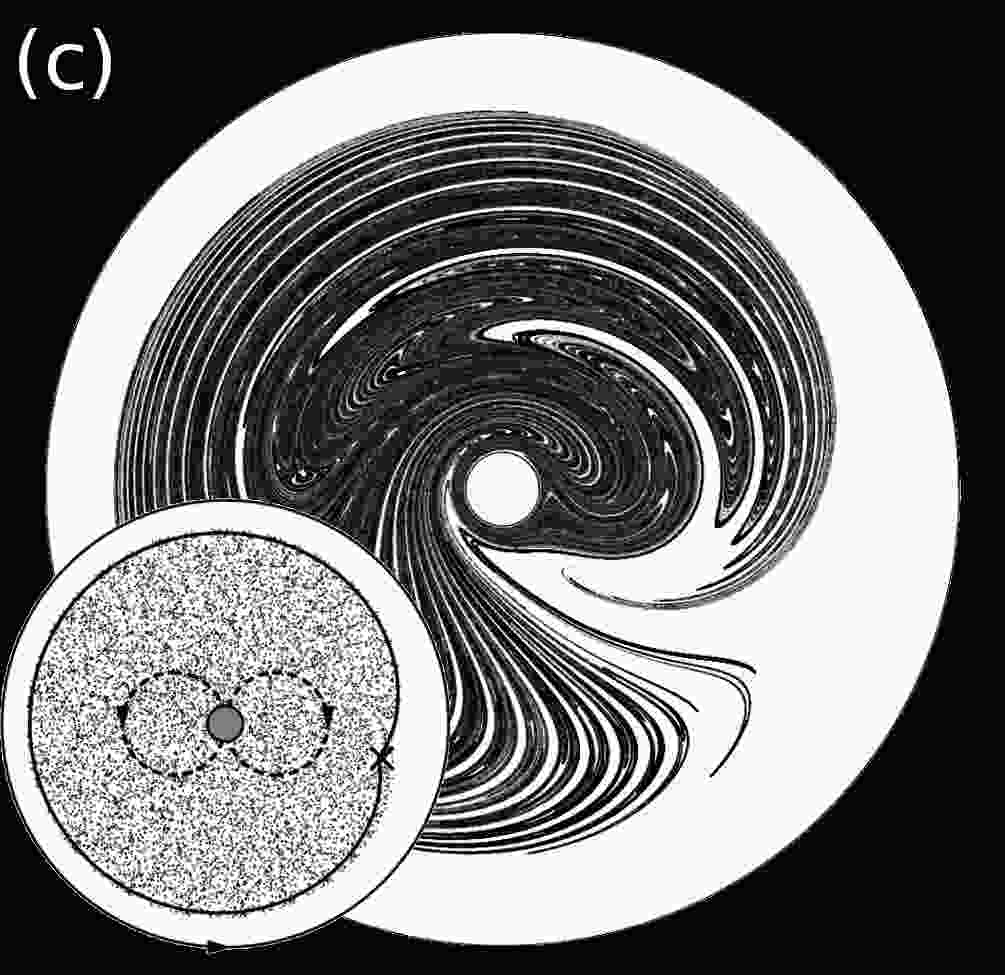}
\includegraphics[width=0.47\columnwidth, height=0.47\columnwidth]{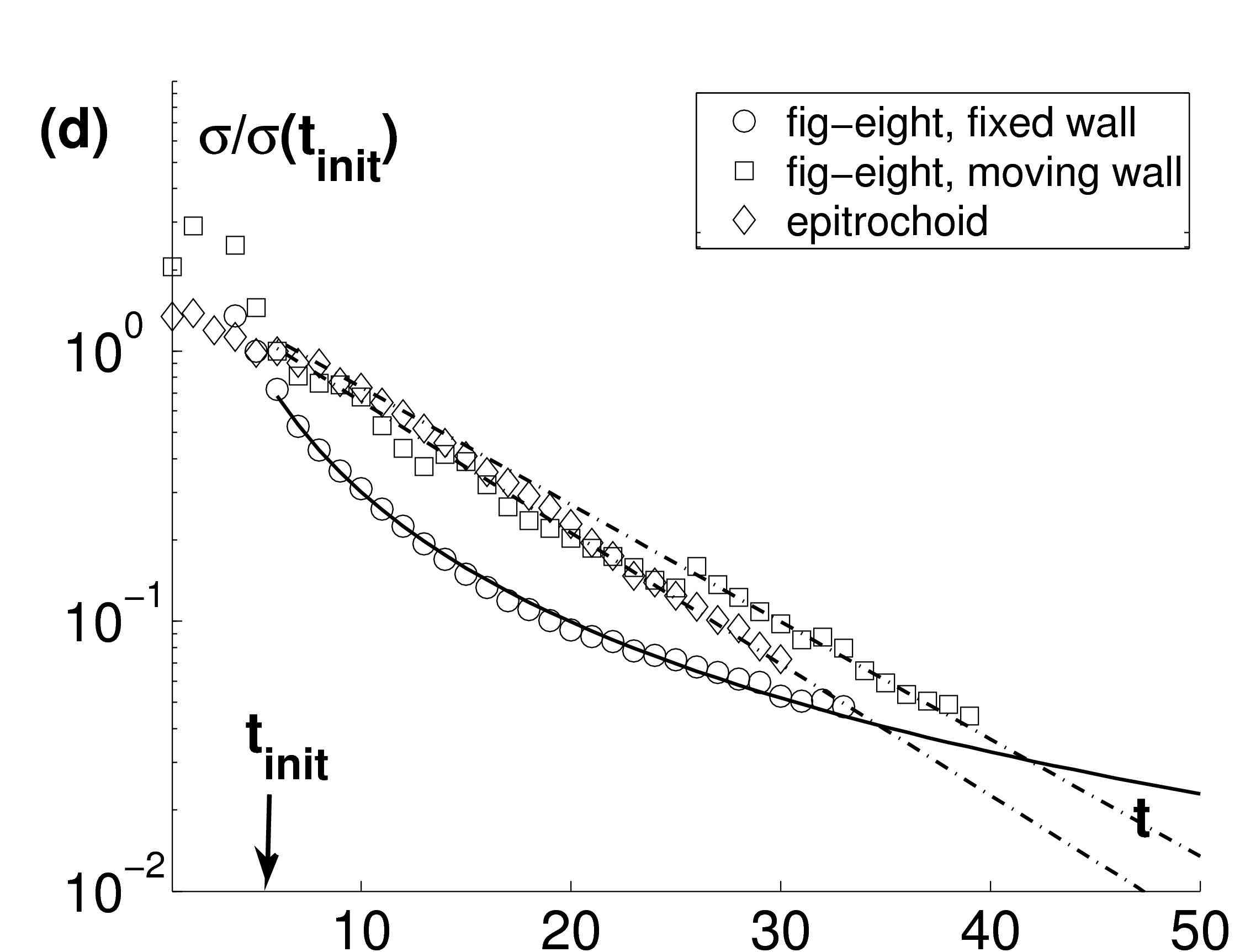}
\end{center}
\vspace{-0.3cm}
\caption{Chaotic mixing in a closed vessel. (a)--(b) Experiments: a
  rod moves periodically following (a) a figure-eight and (b) an
  epitrochoid path.  (c) Numerical simulation of the figure-eight path
  with a rotating wall.  Insets: Poincar\'e section (stroboscopic map)
  obtained numerically for the corresponding Stokes flow (dotted line:
  rod path; filled circles: fixed points of interest; solid line:
  associated separatrices). (d) Evolution of the standard
  deviation~$\sigma(C)$ of the concentration, rescaled by its value at
  $t_{init}$, the time at which the mixing pattern emerges, for
  (a)--(c).}
\label{fig:mixer}
\end{figure}

\begin{figure}[t!]
\begin{center}
\includegraphics[width=\columnwidth]{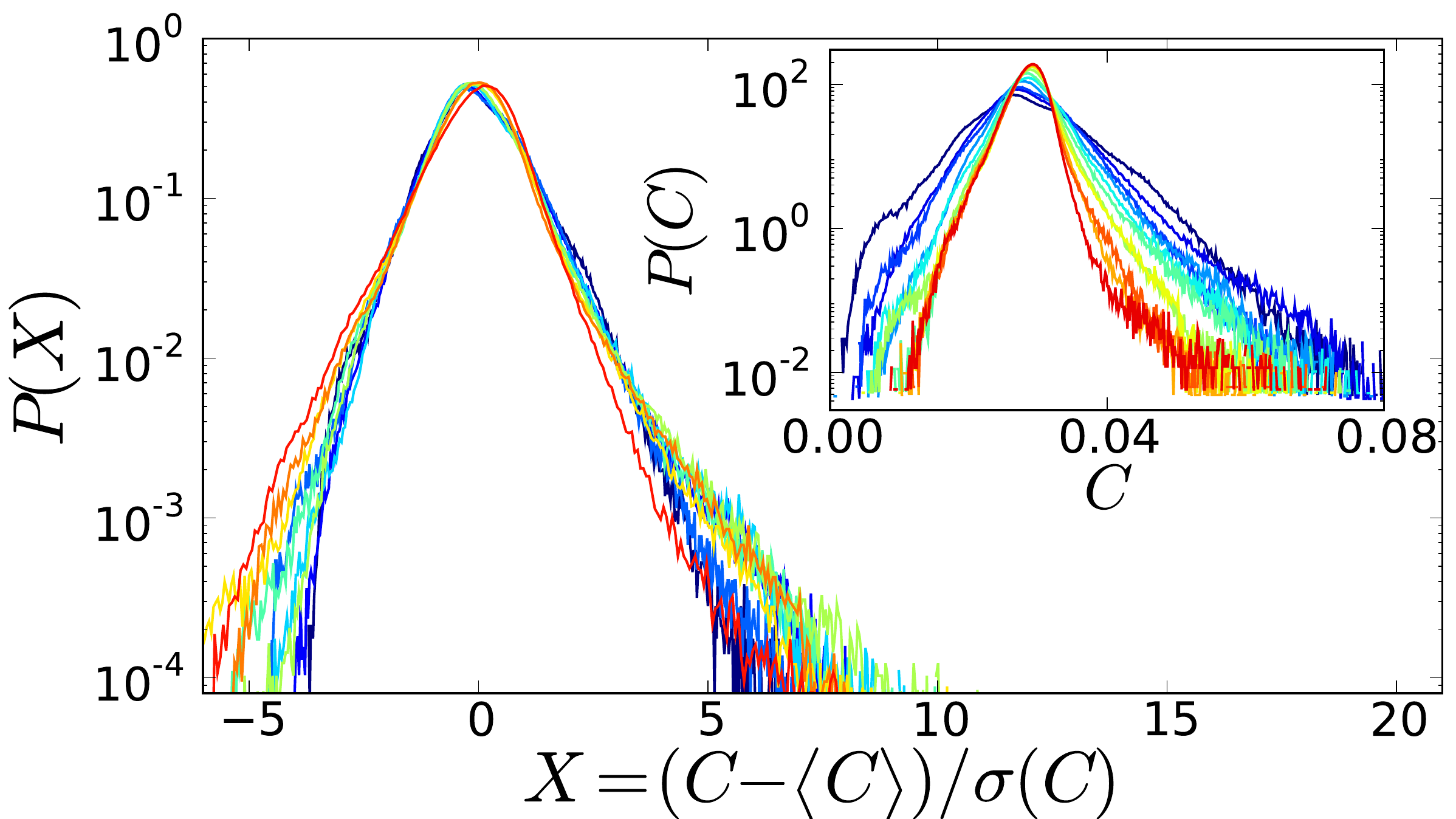}
\vskip 0.5em
\includegraphics[width=\columnwidth]{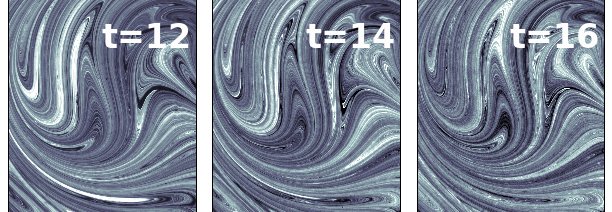}
\end{center}
\caption{[Color online] Experimental evidence of an eigenmode for the
  epitrochoid protocol.  Top: probability distribution function of the
  concentration. Bottom: rescaled concentration field at successive
  periods, showing the converged periodic pattern.}
\vspace{-.5cm}
\label{fig:eigenmode}
\end{figure}

In this Letter, we present a class of time-periodic protocols
exhibiting chaotic advection, but where the chaotic region is shielded
from the influence of the wall. We recover the exponential decay of
the scalar variance (Fig.~\ref{fig:mixer}(d)) and experimental
concentration patterns that converge to a periodic pattern after a few
periods (Fig.~\ref{fig:eigenmode}), as predicted by theories based on
statistics of stretching~\cite{Antonsen1996, Balkovsky1999,
  Thiffeault2003d} or eigenmode analysis of the advection-diffusion
operator~\cite{Pierrehumbert1994, Fereday2002}.  On the contrary, the
convergence to an eigenmode pattern was never observed on experimental
timescales for a chaotic region that extends to the
walls~\cite{Gouillart2007,Gouillart2008}.  The key ingredient is that
a regular region separates the chaotic region from the wall, as in
Fig.~\ref{fig:mixer}(b) and (c). Rotating the wall at a constant
angular velocity is a simple way of creating such a regular layer that
insulates the chaotic region from the wall.  In the present study, we
investigate the case of a slowly moving wall and predict the size of
the regular region, as well as the dynamics near the separatrix
between chaotic and regular regions. We compare results of an
analytical model with numerical simulations.  Finally, we investigate
the influence of the phase portrait on the dynamics of mixing and
deduce an optimum for the rotation velocity.

Let us first present the basic experimental and numerical
observations.  In all protocols, we consider two-dimensional flows
with a single periodically-driven cylindrical stirrer, and an outer
cylindrical wall that is either fixed or rotated at a constant angular
velocity. Experimentally, the rod gently stirs viscous sugar syrup
following either a figure-eight path (Fig.~\ref{fig:mixer}(a)) or an
\emph{epitrochoid} (Fig.~\ref{fig:mixer}(b)), that is a loop with a
smaller inner loop. The fluid viscosity $\nu=5\times 10^{-4}\,
\text{m}^2\cdot\text{s}^{-1}$ together with rod diameter $\ell=16
\,\text{mm}$ and stirring velocity $U=2~\text{cm}\cdot\text{s}^{-1}$
yield a Reynolds number $Re = U\ell/\nu \simeq 0.6$, consistent with a
Stokes-flow regime. A spot of low-diffusivity dye (India ink diluted
in sugar syrup) is injected at the surface of the fluid, and we follow
the evolution of the dye concentration field during the mixing process
(see Fig.~\ref{fig:mixer}).  The concentration field is measured
through the transparent bottom of the vessel using a $12$-bit CCD
camera at resolution $2000 \times 2000$. Since our experimental set-up
does not allow rotation of the vessel wall, we resort to numerical
simulations of Stokes flow~\cite{MattFinn2001} for integrating
Lagrangian trajectories of the figure-eight protocol with a rotating
wall. A scalar concentration field is obtained by coarse-graining the
positions of a large number ($4\times 10^6$) of trajectories, all
initialized inside the same small spot. Figure~\ref{fig:mixer}(c)
shows a numerically-computed concentration field for the case of a
rotating wall.

In all cases, the scalar blob, initially released in the center of the
vessel, is stretched and folded into thin filaments, resulting in a
complicated pattern that expands with time and gradually fills the
chaotic region. The extent of the chaotic region can be seen in the
numerically-computed Poincar\'e sections in the insets of
Fig.~\ref{fig:mixer}(a--c).  For the figure-eight protocol with fixed
walls (Fig.~\ref{fig:mixer}(a)), the chaotic region spans the entire
vessel.  By contrast, the other protocols force fluid particles to
follow regular closed trajectories in the vicinity of the wall, either
because of the rotation of the wall itself (Fig.~\ref{fig:mixer}(c)),
or because the epitrochoid protocol imposes a net rotational motion of
the fluid elements close to the wall (Fig.~\ref{fig:mixer}(b)).  A
regular (non-chaotic) unmixed region thus separates the walls from the
chaotic region in Fig.~\ref{fig:mixer}(b--c). This is the key factor
responsible for the different evolutions of the standard
deviation~$\sigma(C)$ of the concentration field shown in
Fig.~\ref{fig:mixer}(d). Statistics of the concentration field are
measured in a large rectangle inside the chaotic region. In the case
of the figure-eight protocol with fixed wall, the standard deviation
decays algebraically, whereas it decays exponentially in the two other
cases.  For the epitrochoid protocol, we have experimental evidences
that the concentration fields rapidly converges to an eigenmode (see
Fig.~\ref{fig:eigenmode}): the probability distribution of the
concentration and the concentration pattern itself rapidly become
invariant. This periodic pattern is associated with the
slowest-decaying Floquet eigenmode of the advection-diffusion
operator~\cite{Pierrehumbert1994}.  By contrast, we described
in~\cite{Gouillart2007,Gouillart2008} how no permanent pattern can be
attained in the case of a fully chaotic region, because an unmixed
pool of fluid close to the wall slowly leaks into the center of the
fluid domain.

In~\cite{Gouillart2007,Gouillart2008}, we showed how the algebraic
variance decay is related to the algebraic dynamics of the flow in the
vicinity of a parabolic fixed point located at the wall. In the
following, we will show how the new phase portrait obtained in the
presence of a net rotation within the vessel is characterized by a
regular region near the wall of typical width~$\gap$.  The
convergence of fluid particles to this region is exponentially fast,
and we recover an exponential decrease of the variance.  We focus on
the figure-eight rod-stirring protocol with a moving wall, where we
can tune the rotation rate of the wall independently from the rod
motion. In the limit of slowly moving walls, incompressibility and the
no-slip boundary condition suggest writing the flow near the wall as a
simple map,
\begin{equation}
  \xcp = \xc + \Uc + \A(\xc)\yc, \qquad
  \ycp = \yc - \tfrac12\A'(\xc)\yc^2,
\label{eq:nearwall}%
\end{equation}
to leading order in~$\yc$.  Here~$(\xc(\time),\yc(\time))$ is the
position of a fluid particle at the beginning of a time
interval,~$(\xcp,\ycp)=(\xc(\time+\T),\yc(\time+\T))$ its new position
after one stirring period $T$, and we have supposed that the fluid is
incompressible.  The angle~$\xc$ is measured counterclockwise along
the wall,~$0\le\yc\ll1$ measures the distance from the wall, and
$\Uc/T$ is the wall angular velocity.  The circular container is
assumed to have unit radius.  We can regard~$\A(\xc)$ as the
contribution of the rod motion to the velocity field near the wall.
As such, $\A(\xc)$ does not vary significantly when $\Uc$ changes, as
long as~$\Uc$ remains small (this has been confirmed numerically).
$\A(\xc)$ is positive where the motion of the rod and the wall
reinforce each other (\ie\ in the left part of the vessel in
Fig.~\ref{fig:mixer}(c)) and negative where there is opposition
between the wall and the rod (right part of the vessel). In
particular, there exists a fixed point of the map~\eqref{eq:nearwall}
(a period-1 point of the flow) given by
\begin{equation}
  \A'(\xcfp)=0\,, \qquad \ycfp=-{\Uc}/{\A(\xcfp)}
\label{eq:fixedpoint}%
\end{equation}%
where the rotating wall balances exactly the flow induced by the moving
rod. This fixed point is at a distance
\begin{equation}
  \gaps(\Uc)={\Uc}/{\lvert\A(\xcfp)\rvert}    
\label{eq:ee}
\end{equation}
from the wall and is pictured as a small filled disk in
Fig.~\ref{fig:mixer}(c).
\begin{figure}[t!]
\includegraphics[width=0.52\columnwidth, height=0.5\columnwidth]{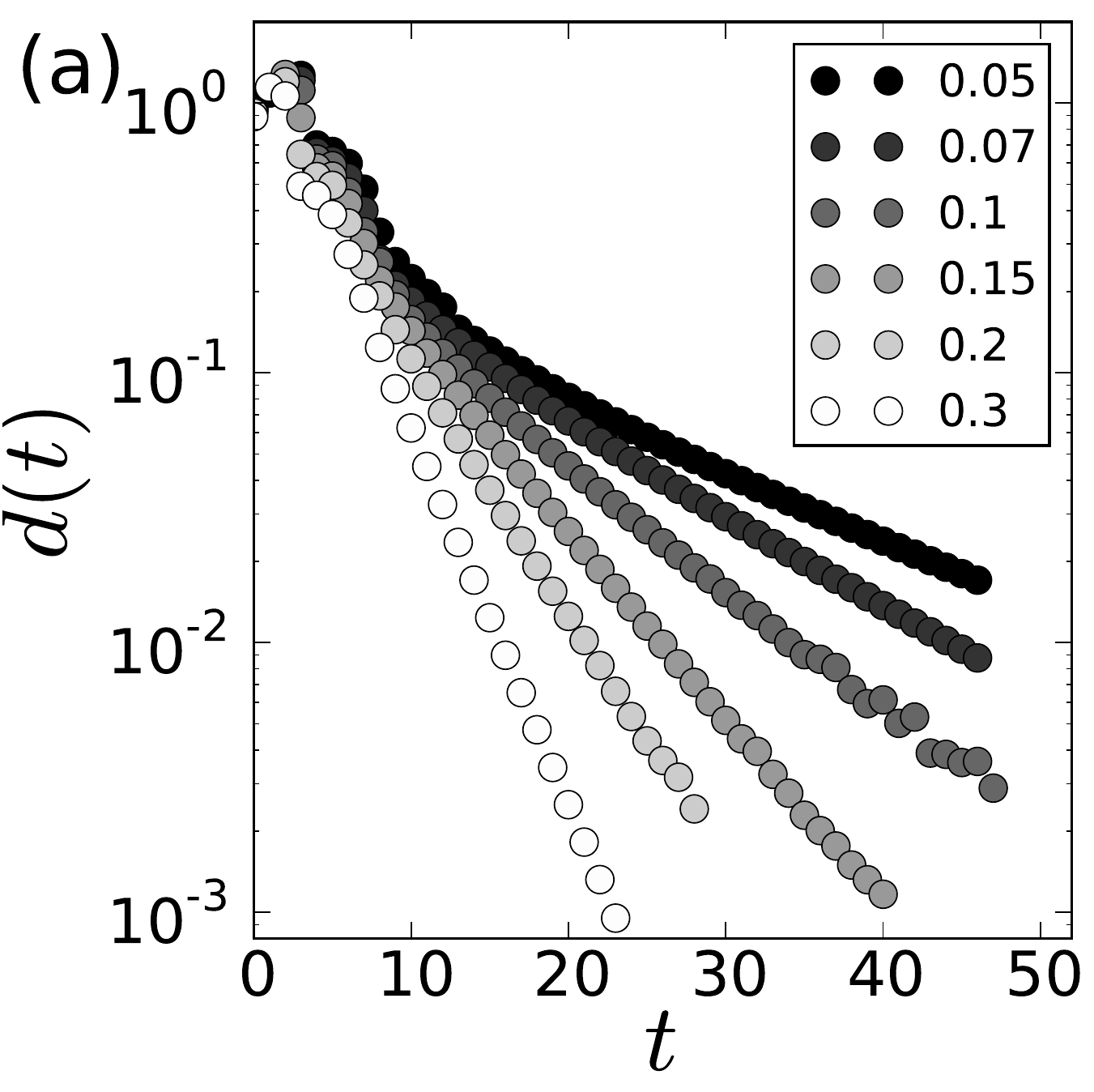}
\includegraphics[width=0.46\columnwidth, height=0.5\columnwidth]{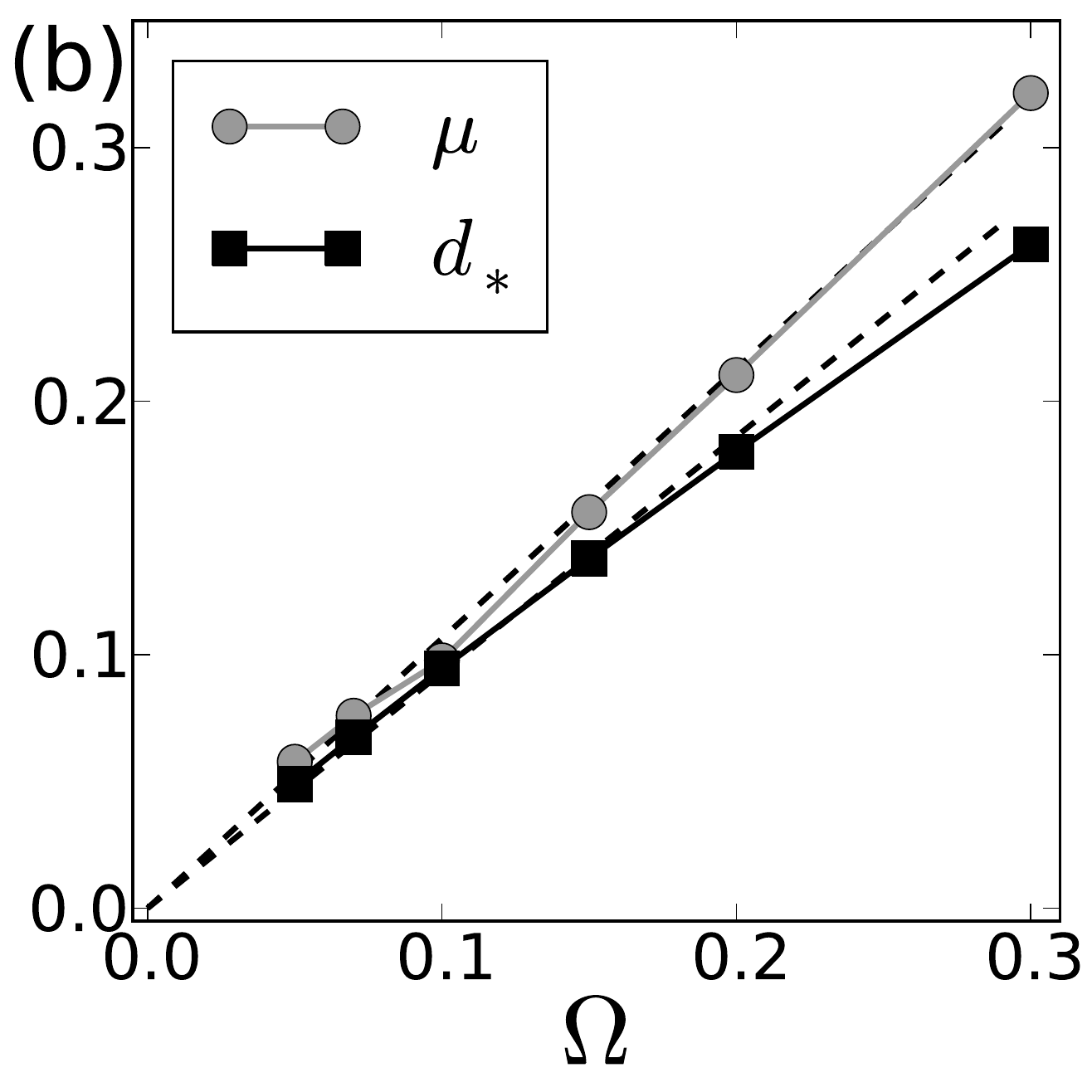}
\caption{(a) The distance~$\gap(\time)$ from the edge of the central mixed
  region to the wall, for different values of $\Uc$.
(b) The exponential decay rate $\decfp(\Uc)$
of $\gap(\time)$ and the distance $\gaps(\Uc)$ separating the hyperbolic
fixed point from the wall, measured from simulations (solid lines) and
from Eqs.~\eqref{eq:ee} and~\eqref{eq:ev} (dashed
lines).}
\vspace{-0.5cm}
\label{fig:fig3}
\end{figure}
The linearization matrix for the dynamics near the fixed point has
eigenvalues $1 \pm \decfp$ with
\begin{equation}
  \decfp = \Uc\,\sqrt{-{\A''(\xcfp)}/{2\A(\xcfp)}}
\label{eq:ev}
\end{equation}
where the argument in the square root is non-negative since~$\A(\xcfp)<0$
and~$\A''(\xcfp)\ge0$. This is a hyperbolic fixed point and the approach
along its stable manifold (the separatrix plotted in
Fig.~\ref{fig:mixer}(c)) is given by
\begin{equation}
  \yc(\time) \sim \yc(0) \,\exp(-\decfp\,\time/\T)
  \label{eq:asymhypo}
\end{equation}
to first order in $\Uc$. The approach to the fixed point is
exponential, at a rate proportional to the speed of rotation of the
wall. For comparison, it was shown
in~\cite{Gouillart2007,Gouillart2008} that the approach to the wall
scales as $1/\time$ for $\Uc=0$.  Figure~\ref{fig:fig3}(c) displays
the evolution of $\gap(t)$, the distance separating the scalar blob
from the fixed point, for simulations with different $\Uc$. One
clearly observes the predicted exponential approach, with a slower
rate for smaller~$\Uc$ (eventually leading to an algebraic decay for
$\Uc=0$). In Fig.~\ref{fig:fig3}(b) we show the results for
$\gaps(\Uc)$ and $\decfp(\Uc)$ defined in Eqs.~\eqref{eq:ee}
and~\eqref{eq:ev}, as measured in numerical simulations. We observe a
linear evolution of $\gaps(\Uc)$ and $\decfp(\Uc)$ with $\Uc$ that
agrees quantitatively with Eqs.~\eqref{eq:ee} and~\eqref{eq:ev}.  Even
a slow rotation of the wall ($\Uc=0.3$ corresponds to a whole rotation
of the vessel in 21 periods of the rod motion) decreases the radius of
the chaotic region by $25\%$.  Altogether, the agreement of the above
analysis with numerical simulations is excellent.

\begin{figure}[t!]
\includegraphics[width=0.49\columnwidth]{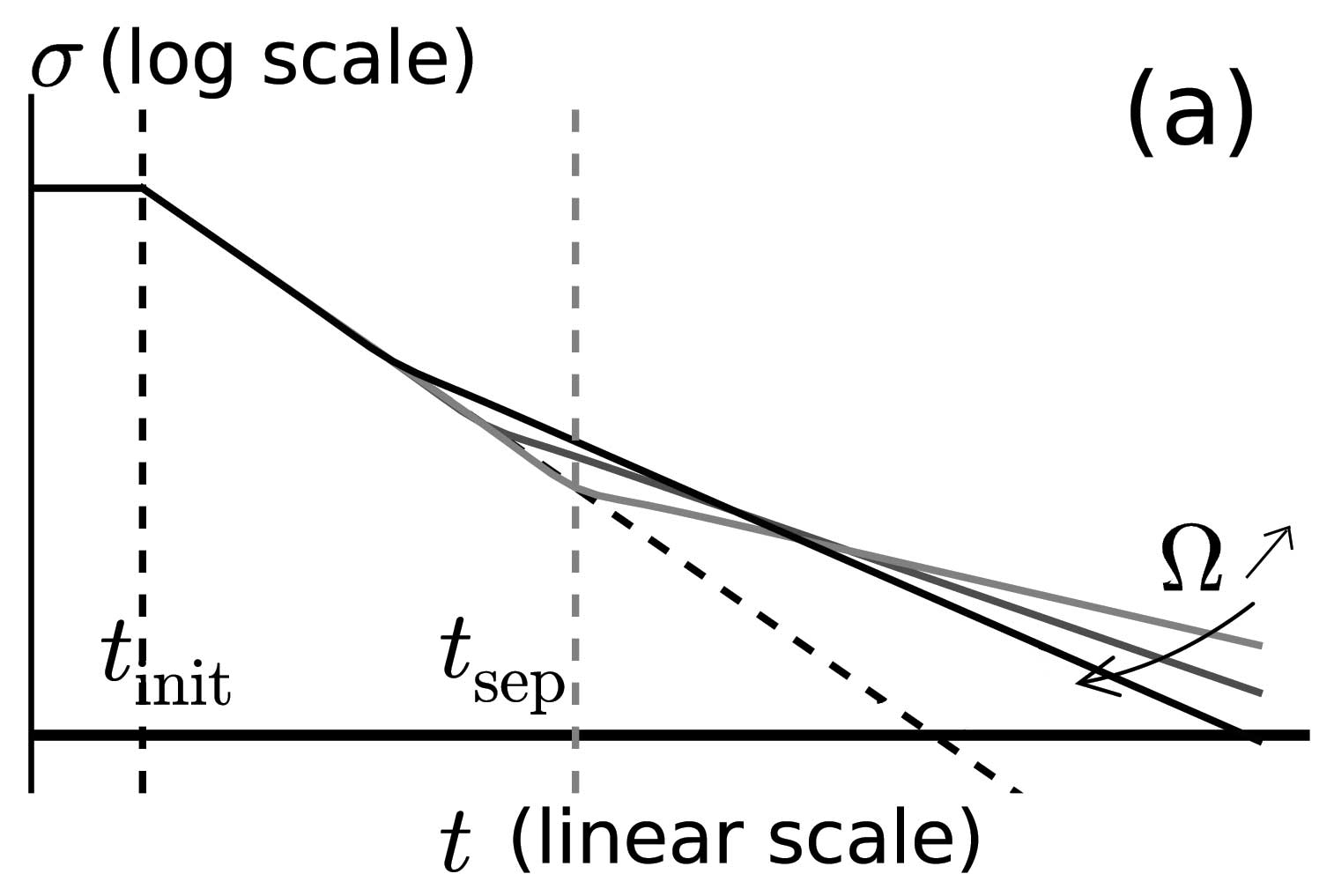}
\includegraphics[width=0.49\columnwidth]{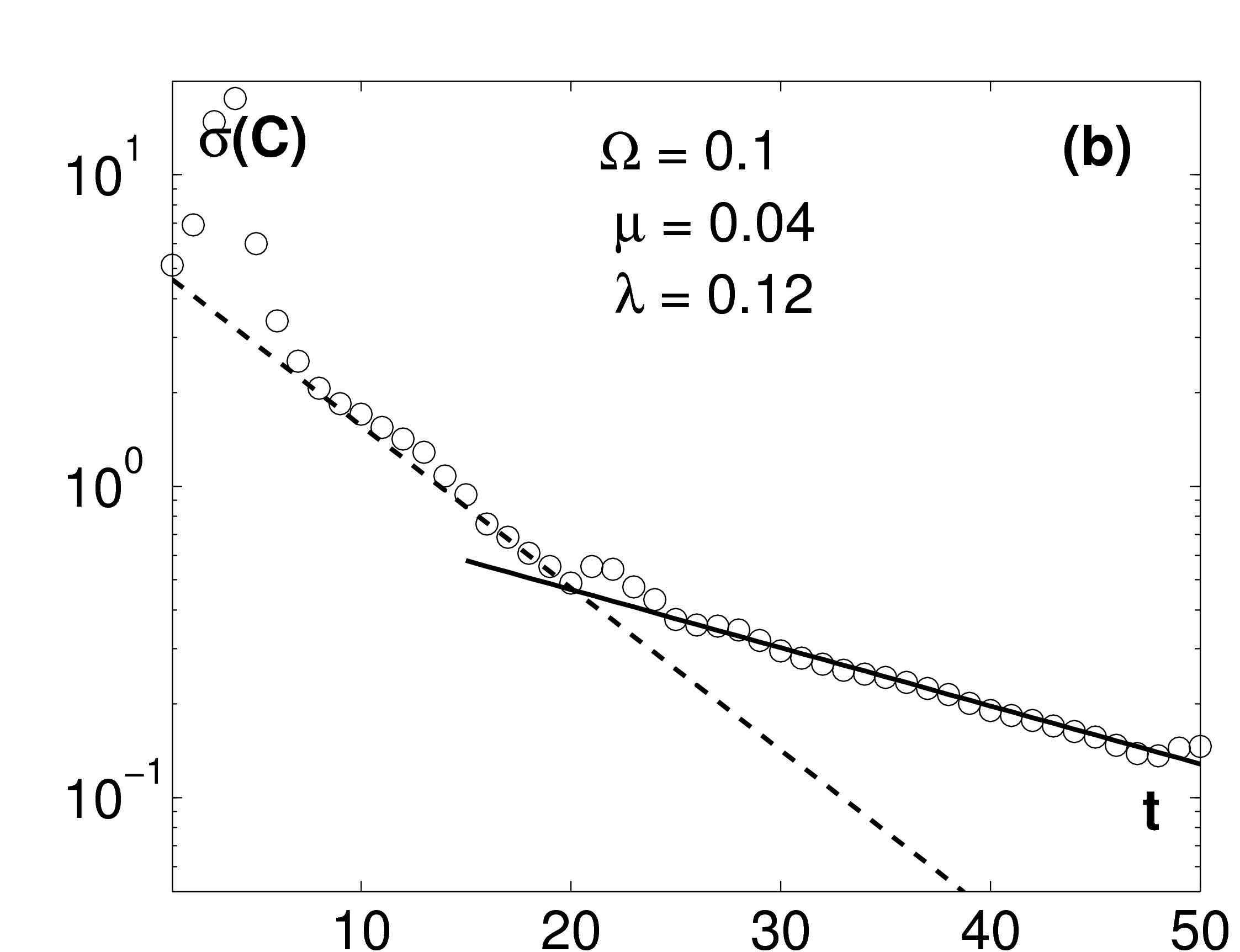}
\includegraphics[width=0.49\columnwidth]{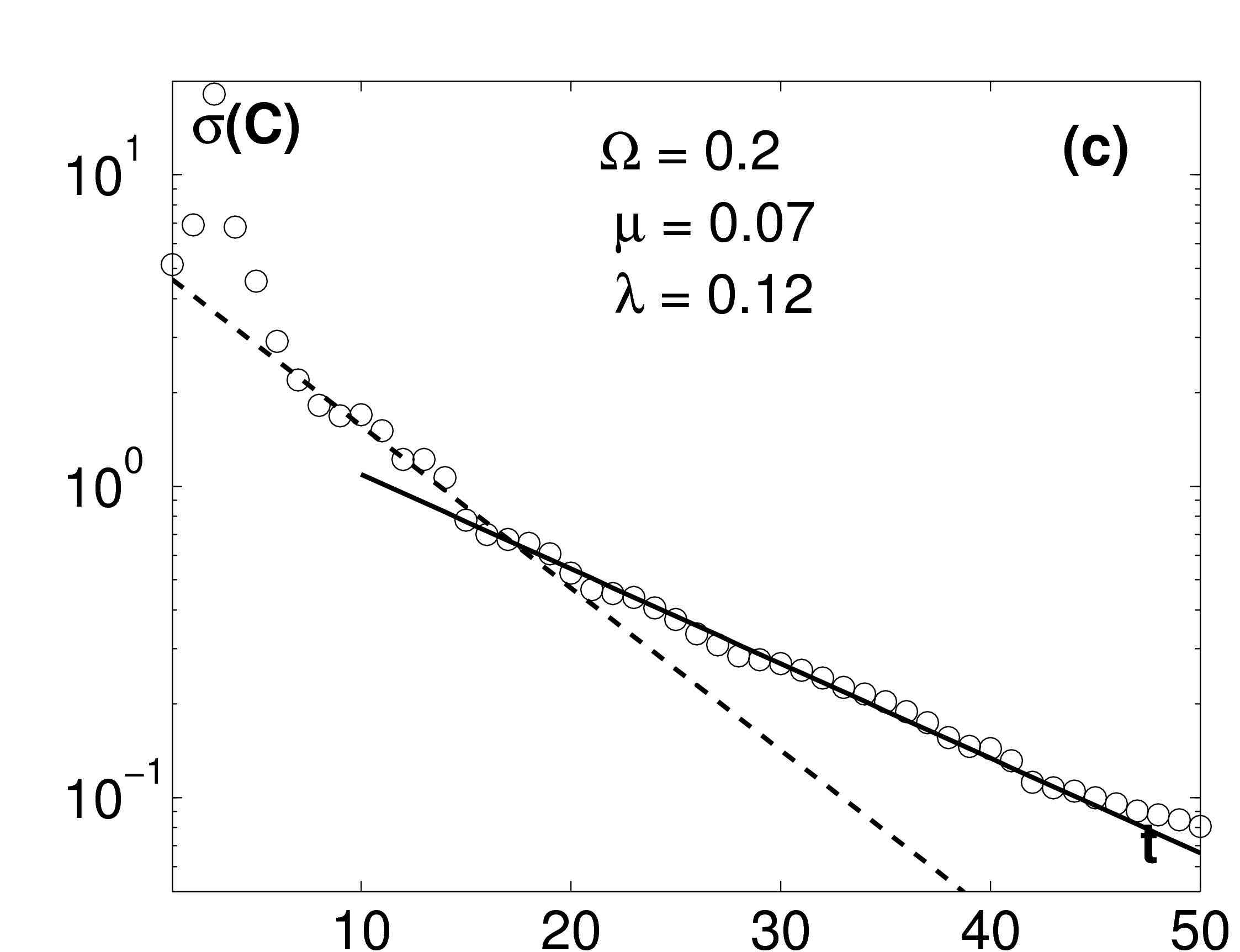}
\includegraphics[width=0.49\columnwidth]{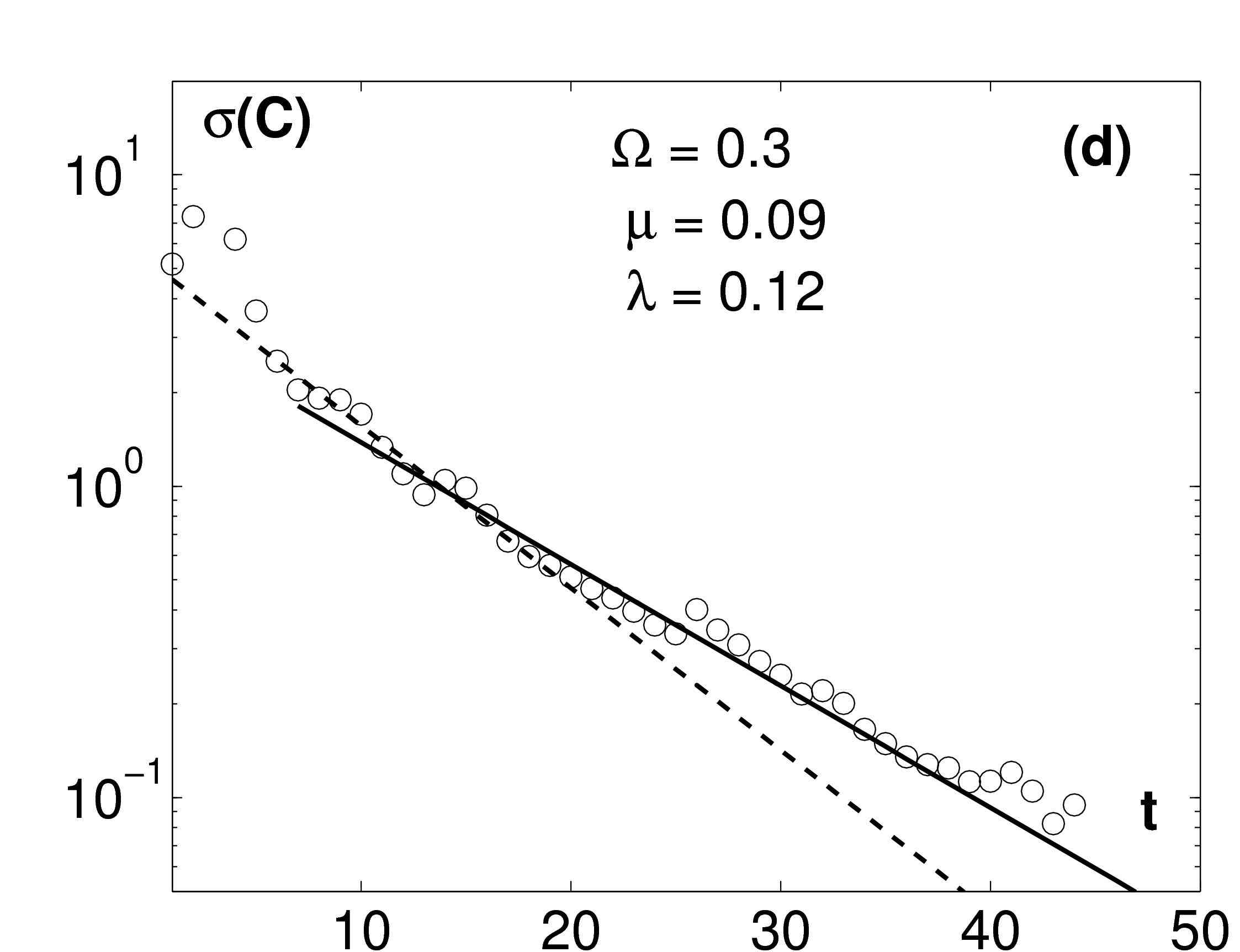}
\caption{Evolution of the standard deviation~$\sigma(C)$. (a) Sketch
  of the decay of~$\sigma(C)$ according to the proposed scenario;
  (b)--(d) Actual decay measured in simulations of the figure-eight
  protocol, for three different wall velocities~$\Uc$; the continuous
  (resp. dash) line shows an exponential decay with rate $\mu/2$
  (resp. $\lambda/2$).}
\label{fig:variance}
\vspace{-0.5cm}
\end{figure}

The different rates of approach are reflected in a different
time-evolution for the scalar variance.  To show this, we have
simulated the mixing of a scalar blob for different wall velocities
($\Uc$ varying from $0.05$ to $0.4$). The evolution of the normalized
standard deviation of the scalar field is measured inside a central
rectangle, and plotted in
Figs.~\ref{fig:variance}(b--d). Fig.~\ref{fig:variance}(a) offers a
simplified sketch of the observed evolution. For a few initial periods
($\time<\tin$), the mixing pattern emerges and diffusion is not yet
efficient.  (Figs.~\ref{fig:variance}(b--d) show a spurious initial
`spike' since the concentration is not measured in the entire domain.)
For times $\time>\tin$ we observe a rapid exponential decrease
of~$\sigma(C)$ for all $\Uc$.  After a time~$\tsep$, we observe a
transition to a slower exponential regime with a rate $\decvar$ that
increases with $\Uc$.  The transition time $\tsep$ decreases
when~$\Uc$ increases (see
Figs.~\ref{fig:variance}(b--d)). Increasing~$\Uc$ further leads to the
collapse of this second exponential regime on the continuation of the
initial exponential regime (see Fig.~\ref{fig:variance}(d)).

We now propose a scenario that accounts for these observations.
Between $\tin$ and $\tsep$, diffusion starts to smear out dye
filaments that are elongated by the rod motion. The decay of the
variance stems from the distribution of stretching experienced by
fluid particles in the bulk, as described in~\cite{Antonsen1996}.  Since
the influence of the wall on the velocity field is weak in the region
spanned by the rod's path, the same decay regime is observed for
different values of~$\Uc$. At the time~$\tsep$ when some filaments
arrive in the vicinity of the hyperbolic point and its separatrix
(visible as a `knee' in Fig.~\ref{fig:fig3}), fluid particles that have
stayed most of the time in the core of the chaotic region have all
been stretched significantly, and most of the initial variance
corresponding to such fluid particles has been exhausted. However,
there exists outside the central mixing pattern a pool of poorly
stretched fluid delineated by the separatrix. For~$\time>\tsep$, the
decay of the distance~$\gap(\time)$ to this separatrix is governed by
the exponential dynamics of Eq.~\eqref{eq:asymhypo}, and because of
area-preservation an unmixed strip of width
$\Delta(\time)\simeq\gap(\time) - \gap(\time+\T) \simeq
-\T\,\dotgap(\time)$ must enter the central mixing region along the
unstable manifold of the fixed point (see the white `tongue' in
Fig.~\ref{fig:mixer}(c)). The strip is then thinned to the Batchelor
length $\ell = \sqrt{\kappa/\lambda}$, where~$\kappa$ is the molecular
diffusivity and $\lambda$ is an average stretching rate inside the
central mixing
region~\cite{Batchelor1959,Balkovsky1999,ThiffeaultAosta2004}, and
finally wiped out by diffusion.

The fluctuations of the mixing pattern in Fig.~\ref{fig:mixer}(c)
are dominated by the white strips injected most
recently, as well as a few folds of strips injected
previously---portions of the strips that are stretched but not folded
reach $\ell$ faster.  We therefore estimate that the total white area
inside the bulk is the dominant contribution to the variance, and that
it is proportional to $|\dotgap(\time)| \sim
\exp(-\decfp\,\time/\T)$. We observe indeed a decay of the variance at
a rate close to $\decfp$ for small values of $\Uc$
(Fig.~\ref{fig:variance}(b--c)), consistent with a rate of mixing
dominated by the rate of approach to the hyperbolic fixed point. The
central mixing process is potentially more efficient
($\lambda>\decfp$), but it is starved by the boundaries. When the
rotation of the wall is increased so that $\decfp$ becomes of the same
order as $\lambda$ or greater, stretching in the bulk is as slow as
that near the separatrix (or slower), so that the decay rate
saturates at the same value $\lambda$ as during the initial decay
phase.

In summary, we have proposed a design principle for mixing protocols
that protect the chaotic mixing region from the slowdown due to
no-slip walls. Creating a net rotation of the fluid in the vicinity of
the wall results in a mixed phase space with a regular region that
insulates the central chaotic region from the wall.  In this case, we
retrieve an exponential decay of scalar variance.  For the case of a
rotating vessel wall, rotating faster may increase the mixing
efficiency.  However, we have seen that the variance decay rate is
always bounded by that of the bulk, and that when increasing $\Uc$ the
total mixing region is smaller, so the influence of the separatrix is
felt earlier. An understanging of this mechanism is a first step
towards the design of efficient mixers. Depending on the application,
it may for instance be appropriate to rotate the wall in such a way
that $\decfp(\Uc)\simeq\lambda$, to match the rate of stretching in
the bulk to that near the outer hyperbolic point.

From a practical point of view, we note that for the epitrochoid
protocol the rod itself generates the net rotation of the fluid
elements, which qualitatively induces similar effects as above, but
without the need for a rotating wall. Obviously, the biggest drawback
of the design principle presented here---to insulate the chaotic
region from the walls--- is the presence of an unmixed region near the
wall. But this may be unimportant in applications where one can
selectively sample the mixture, for instance in pharmaceuticals.  In
any case, there is no other known practical way to avoid a
wall-induced slowdown of the variance decay.  Finally, from a more
fundamental point of view it is intriguing that in all cases,
including the epitrochoid, the decay rate inside the mixing region is
always rather small, as compared for instance to that of a baker's
map. The origin of this slow mixing---the distribution of hyperbolic
points, parabolic points, etc.~\cite{Antonsen1996, Balkovsky1999,
  Thiffeault2003d, Gouillart2008}---is a key issue to investigate in
the future.

The authors thank Matthew D. Finn for the use of his computer code for
simulating viscous flows, as well as C\'ecile Gasquet and Vincent Padilla
for technical assistance. J-LT was also partially supported by the
Division of Mathematical Sciences of the US National Science Foundation,
under grant DMS-0806821.

\vspace{-1.5em}


\begin{thebibliography}{22}
\expandafter\ifx\csname natexlab\endcsname\relax\def\natexlab#1{#1}\fi
\expandafter\ifx\csname bibnamefont\endcsname\relax
  \def\bibnamefont#1{#1}\fi
\expandafter\ifx\csname bibfnamefont\endcsname\relax
  \def\bibfnamefont#1{#1}\fi
\expandafter\ifx\csname citenamefont\endcsname\relax
  \def\citenamefont#1{#1}\fi
\expandafter\ifx\csname url\endcsname\relax
  \def\url#1{\texttt{#1}}\fi
\expandafter\ifx\csname urlprefix\endcsname\relax\def\urlprefix{URL }\fi
\providecommand{\bibinfo}[2]{#2}
\providecommand{\eprint}[2][]{\url{#2}}

\bibitem[{\citenamefont{Aref}(1984)}]{Aref1984}
\bibinfo{author}{\bibfnamefont{H.}~\bibnamefont{Aref}}, \bibinfo{journal}{J.
  Fluid Mech.} \textbf{\bibinfo{volume}{143}}, \bibinfo{pages}{1}
  (\bibinfo{year}{1984}).

\bibitem[{\citenamefont{Ottino}(1989)}]{Ottino}
\bibinfo{author}{\bibfnamefont{J.~M.} \bibnamefont{Ottino}},
  \emph{\bibinfo{title}{The Kinematics of Mixing: Stretching, Chaos, and
  Transport}} (\bibinfo{publisher}{Cambridge University Press},
  \bibinfo{address}{Cambridge, U.K.}, \bibinfo{year}{1989}).

\bibitem[{\citenamefont{Pikovsky and Popovych}(2003)}]{Pikovsky2003}
\bibinfo{author}{\bibfnamefont{A.}~\bibnamefont{Pikovsky}} \bibnamefont{and}
  \bibinfo{author}{\bibfnamefont{O.}~\bibnamefont{Popovych}},
  \bibinfo{journal}{Europhys. Lett.} \textbf{\bibinfo{volume}{61}},
  \bibinfo{pages}{625} (\bibinfo{year}{2003}).



\bibitem[{\citenamefont{groupedtheory}(0000)}]{groupedtheory}
\bibinfo{author}{\bibfnamefont{M.}~\bibnamefont{Chertkov}} \bibnamefont{and}
  \bibinfo{author}{\bibfnamefont{V.}~\bibnamefont{Lebedev}},
  \bibinfo{journal}{Phys. Rev. Lett.} \textbf{\bibinfo{volume}{90}},
  \bibinfo{pages}{034501} (\bibinfo{year}{2003});
\bibinfo{author}{\bibfnamefont{V.~V.} \bibnamefont{Lebedev}} \bibnamefont{and}
  \bibinfo{author}{\bibfnamefont{K.~S.} \bibnamefont{Turitsyn}},
  \bibinfo{journal}{Phys. Rev. E} \textbf{\bibinfo{volume}{69}},
  \bibinfo{pages}{036301} (\bibinfo{year}{2004});
\bibinfo{author}{\bibfnamefont{H.}~\bibnamefont{Salman}} \bibnamefont{and}
  \bibinfo{author}{\bibfnamefont{P.~H.} \bibnamefont{Haynes}},
  \bibinfo{journal}{Phys. Fluids} \textbf{\bibinfo{volume}{19}},
  \bibinfo{pages}{067101} (\bibinfo{year}{2007});
\bibinfo{author}{\bibfnamefont{O.}~\bibnamefont{Popovych}},
  \bibinfo{author}{\bibfnamefont{A.}~\bibnamefont{Pikovsky}}, \bibnamefont{and}
  \bibinfo{author}{\bibfnamefont{B.}~\bibnamefont{Eckhardt}},
  \bibinfo{journal}{Phys. Rev. E} \textbf{\bibinfo{volume}{75}},
  \bibinfo{pages}{036308} (\bibinfo{year}{2007}).


\bibitem[{\citenamefont{groupednumerics}(0000)}]{groupednumerics}
\bibinfo{author}{\bibfnamefont{S.~C.} \bibnamefont{Jana}},
  \bibinfo{author}{\bibfnamefont{G.}~\bibnamefont{Metcalfe}}, \bibnamefont{and}
  \bibinfo{author}{\bibfnamefont{J.~M.} \bibnamefont{Ottino}},
  \bibinfo{journal}{J. Fluid Mech.} \textbf{\bibinfo{volume}{269}},
  \bibinfo{pages}{199} (\bibinfo{year}{1994});
\bibinfo{author}{\bibfnamefont{G.}~\bibnamefont{Boffetta}},
  \bibinfo{author}{\bibnamefont{{F. De Lillo}}}, \bibnamefont{and}
  \bibinfo{author}{\bibnamefont{{A. Mazzino}}} (\bibinfo{year}{2008}),
  \eprint{arXiv:0811.4519};
\bibinfo{author}{\bibfnamefont{K.}~\bibnamefont{{El Omari}}} \bibnamefont{and}
  \bibinfo{author}{\bibfnamefont{Y.}~\bibnamefont{{Le Guer}}},
  \bibinfo{journal}{Comput. Therm. Sci.} \textbf{\bibinfo{volume}{1}},
  \bibinfo{pages}{55} (\bibinfo{year}{2009}).


\bibitem[{\citenamefont{Gouillart et~al.}(2007)\citenamefont{Gouillart, Kuncio,
  Dauchot, Dubrulle, Roux, and Thiffeault}}]{Gouillart2007}
\bibinfo{author}{\bibfnamefont{E.}~\bibnamefont{Gouillart}},
  \bibinfo{author}{\bibfnamefont{N.}~\bibnamefont{Kuncio}},
  \bibinfo{author}{\bibfnamefont{O.}~\bibnamefont{Dauchot}},
  \bibinfo{author}{\bibfnamefont{B.}~\bibnamefont{Dubrulle}},
  \bibinfo{author}{\bibfnamefont{S.}~\bibnamefont{Roux}}, \bibnamefont{and}
  \bibinfo{author}{\bibfnamefont{J.-L.} \bibnamefont{Thiffeault}},
  \bibinfo{journal}{Phys. Rev. Lett.} \textbf{\bibinfo{volume}{99}},
  \bibinfo{pages}{114501} (\bibinfo{year}{2007}).

\bibitem[{\citenamefont{Gouillart et~al.}(2008)\citenamefont{Gouillart,
  Dauchot, Dubrulle, Roux, and Thiffeault}}]{Gouillart2008}
\bibinfo{author}{\bibfnamefont{E.}~\bibnamefont{Gouillart}},
  \bibinfo{author}{\bibfnamefont{O.}~\bibnamefont{Dauchot}},
  \bibinfo{author}{\bibfnamefont{B.}~\bibnamefont{Dubrulle}},
  \bibinfo{author}{\bibfnamefont{S.}~\bibnamefont{Roux}}, \bibnamefont{and}
  \bibinfo{author}{\bibfnamefont{J.-L.} \bibnamefont{Thiffeault}},
  \bibinfo{journal}{Phys. Rev. E} \textbf{\bibinfo{volume}{78}},
  \bibinfo{pages}{026211} (\bibinfo{year}{2008}).

\bibitem[{\citenamefont{Gouillart et~al.}(2009)\citenamefont{Gouillart,
  Dauchot, Thiffeault, and Roux}}]{Gouillart2009}
\bibinfo{author}{\bibfnamefont{E.}~\bibnamefont{Gouillart}},
  \bibinfo{author}{\bibfnamefont{O.}~\bibnamefont{Dauchot}},
  \bibinfo{author}{\bibfnamefont{J.-L.} \bibnamefont{Thiffeault}},
  \bibnamefont{and} \bibinfo{author}{\bibfnamefont{S.}~\bibnamefont{Roux}},
  \bibinfo{journal}{Phys. Fluids} \textbf{\bibinfo{volume}{21}},
  \bibinfo{pages}{022603} (\bibinfo{year}{2009}).

\bibitem[{\citenamefont{{Antonsen, Jr.} et~al.}(1996)\citenamefont{{Antonsen,
  Jr.}, Fan, Ott, and Garcia-Lopez}}]{Antonsen1996}
\bibinfo{author}{\bibfnamefont{T.~M.} \bibnamefont{{Antonsen, Jr.}}},
  \bibinfo{author}{\bibfnamefont{Z.}~\bibnamefont{Fan}},
  \bibinfo{author}{\bibfnamefont{E.}~\bibnamefont{Ott}}, \bibnamefont{and}
  \bibinfo{author}{\bibfnamefont{E.}~\bibnamefont{Garcia-Lopez}},
  \bibinfo{journal}{Phys. Fluids} \textbf{\bibinfo{volume}{8}},
  \bibinfo{pages}{3094} (\bibinfo{year}{1996}).

\bibitem[{\citenamefont{Thiffeault and Childress}(2003)}]{Thiffeault2003d}
\bibinfo{author}{\bibfnamefont{J.-L.} \bibnamefont{Thiffeault}}
  \bibnamefont{and}
  \bibinfo{author}{\bibfnamefont{S.}~\bibnamefont{Childress}},
  \bibinfo{journal}{Chaos} \textbf{\bibinfo{volume}{13}}, \bibinfo{pages}{502}
  (\bibinfo{year}{2003}).

\bibitem[{\citenamefont{Balkovsky and Fouxon}(1999)}]{Balkovsky1999}
\bibinfo{author}{\bibfnamefont{E.}~\bibnamefont{Balkovsky}} \bibnamefont{and}
  \bibinfo{author}{\bibfnamefont{A.}~\bibnamefont{Fouxon}},
  \bibinfo{journal}{Phys. Rev. E} \textbf{\bibinfo{volume}{60}},
  \bibinfo{pages}{4164} (\bibinfo{year}{1999}).

\bibitem[{\citenamefont{Pierrehumbert}(1994)}]{Pierrehumbert1994}
\bibinfo{author}{\bibfnamefont{R.~T.} \bibnamefont{Pierrehumbert}},
  \bibinfo{journal}{Chaos Solitons Fractals} \textbf{\bibinfo{volume}{4}},
  \bibinfo{pages}{1091} (\bibinfo{year}{1994}).

\bibitem[{\citenamefont{Fereday et~al.}(2002)\citenamefont{Fereday, Haynes,
  Wonhas, and Vassilicos}}]{Fereday2002}
\bibinfo{author}{\bibfnamefont{D.~R.} \bibnamefont{Fereday}},
  \bibinfo{author}{\bibfnamefont{P.~H.} \bibnamefont{Haynes}},
  \bibinfo{author}{\bibfnamefont{A.}~\bibnamefont{Wonhas}}, \bibnamefont{and}
  \bibinfo{author}{\bibfnamefont{J.~C.} \bibnamefont{Vassilicos}},
  \bibinfo{journal}{Phys. Rev. E} \textbf{\bibinfo{volume}{65}},
  \bibinfo{pages}{035301(R)} (\bibinfo{year}{2002}).

\bibitem[{\citenamefont{Finn and Cox}(2001)}]{MattFinn2001}
\bibinfo{author}{\bibfnamefont{M.~D.} \bibnamefont{Finn}} \bibnamefont{and}
  \bibinfo{author}{\bibfnamefont{S.~M.} \bibnamefont{Cox}},
  \bibinfo{journal}{J. Eng. Math.} \textbf{\bibinfo{volume}{41}},
  \bibinfo{pages}{75} (\bibinfo{year}{2001}).

\bibitem[{\citenamefont{Batchelor}(1959)}]{Batchelor1959}
\bibinfo{author}{\bibfnamefont{G.~K.} \bibnamefont{Batchelor}},
  \bibinfo{journal}{J. Fluid Mech.} \textbf{\bibinfo{volume}{5}},
  \bibinfo{pages}{113} (\bibinfo{year}{1959}).

\bibitem[{\citenamefont{Thiffeault}(2008)}]{ThiffeaultAosta2004}
\bibinfo{author}{\bibfnamefont{J.-L.} \bibnamefont{Thiffeault}}, in
  \emph{\bibinfo{booktitle}{Transport and Mixing in Geophysical Flows}}, edited
  by \bibinfo{editor}{\bibfnamefont{J.~B.} \bibnamefont{Weiss}}
  \bibnamefont{and}
  \bibinfo{editor}{\bibfnamefont{A.}~\bibnamefont{Provenzale}}
  (\bibinfo{publisher}{Springer}, \bibinfo{address}{Berlin},
  \bibinfo{year}{2008}), \emph{\bibinfo{series}{Lecture Notes in
      Physics}} \textbf{\bibinfo{volume}{744}}, pp.
  \bibinfo{pages}{3--35}, \eprint{arXiv:nlin/0502011}.

\end{thebibliography}

\end{document}